\documentclass{prop2015v2}
\usepackage[english]{babel}

\def\Anti{\mathbf{Anti}}
\def\Adj{\mathbf{Adj}}
\def\1{\mathbf{1}}
\def\2{\mathbf{2}}
\def\3{\mathbf{3}}
\def\4{\mathbf{4}}
\def\N{\mathbf{N}}
\def\ov{\overline}

\def\Z{\mathbb{Z}}
\def\OR{\Omega\mathcal{R}}
\keywords{String Phenomenology \& Cosmology, Intersecting D-branes, Orbifolds, Deformations, Moduli Stabilisation,  Axions.}
\title{From Stringy Particle Physics to Moduli Stabilisation and Cosmology}
\author[G.Honecker]{Gabriele Honecker\inst{1,}\footnote{
E-mail:~\textsf{Gabriele.Honecker@uni-mainz.de}}}
\address[1]{Institute for Physics (WA THEP) \& Cluster of Excellence PRISMA, Johannes Gutenberg University, D - 55099 Mainz}
\shortauthors{G. Honecker}
\begin{abstract}
  Intersecting D6-branes provide a geometrically intuitive road to stringy particle physics models, where
  D6-branes stuck at orbifold singularities can lead to the stabilisation of deformation moduli, and
  the QCD axion can arise from the open string sector in a very constrained way compared to pure field theory.
  We demonstrate this interplay of different physical features here  through an explicit model.
\end{abstract}
\shortabstract
\begin{document}
\maketitle

%
%

\section{Introduction}\label{S:Intro}

While string theory is to date arguably the best framework for a unified description of all fundamental interactions, 
it remains an open question how to reconcile searches for particle physics vacua with cosmological models.
D-branes and open strings in Type II orientifold compactifications provide an attractive test ground for dedicated searches for the Standard Model (SM) or 
Grand Unified Theory (GUT) spectra, with intersecting D6-branes providing especial geometric intuition for $m$-point interactions~\cite{Blumenhagen:2006ci,Ibanez:2012zz}.
On the other hand, cosmological features such as inflation are often attributed to the closed string sector, see e.g.~\cite{Svrcek:2006yi,McAllister:2008hb,Shiu:2015xda} 
and references therein. 
Besides from presenting some improved globally consistent SM vacuum, we will discuss here the occurence of the QCD axion from the open string sector and its mixing with a closed string axion, which constitutes a candidate for the inflaton.

A second confrontation in model building arises from the plethora of Calabi-Yau (CY) manifolds only accessible by supergravity (SUGRA) techniques versus a relatively small
number of toroidal orbifolds, where Conformal Field Theory (CFT) tools are available to derive the low-energy effective field theory beyond leading order.
Moreover, the mathematics of {\it special Lagrangian} three-cycles within CYs required for D6-brane model building has to date been only relatively poorly studied, see e.g.~\cite{Palti:2009bt}. Here, we explore 
the moduli sector close to orbifold singularities and find some flat directions in complex structure moduli space with other directions stabilised at the orbifold point by the existence of D-branes.

\vspace{-4mm}

\section{Stringy Particle Physics}\label{S:ParticlePhysics}

To engineer the Standard Model gauge group $SU(3) \times SU(2) \times U(1)_Y$ with a massless hypercharge, at least three different stacks of D-branes are required,  
with the `standard' embedding of the hypercharge,
\begin{equation}\label{Eq:hypercharge}
U(1)_Y= \frac{U(1)_a}{6}+\frac{U(1)_c+U(1)_d}{2},
\end{equation}
provided by the `Spanish Quiver' ansatz~\cite{Ibanez:2001nd} of four stacks with initial gauge group $U(3)_a \times [U \text{ or } USp](2)_b \times U(1)_c \times U(1)_d$.
Requiring the absence of light exotic matter charged under the strong interactions amounts in particular to demanding the existence of some {\it rigid} cycle on which the $U(3)_a$ 
stack can be wrapped. In the context of toroidal orbifolds, the class of $T^6/(\Z_2 \times \Z_{2M} \times \OR)$ with discrete torsion is of particular interest for $M$ odd~\cite{Forste:2010gw}, 
since there exist exceptional three-cycles stuck at $\Z_2 \times \Z_2$ singularities.

While to our best knowledge, no phenomenologically appealing SM or GUT vacuum has been found for the choice $M=1$~\cite{Blumenhagen:2005tn}, for $M=3'$ with the $\Z_3$ subgroup acting along the full six-torus $T^6=(T^2)^3$ a comprehensive scan showed that only globally consistent Pati-Salam models exist~\cite{Honecker:2012qr}. $M=3$ with the $\Z_3$ subgroup acting only along a four-torus $T^4=(T^2)^2$ offers a much wider variety of combinatorial possibilities of D6-branes due to its free complex structure parameter along the remaining two-torus $T^2$~\cite{Forste:2010gw}. We discuss here model building options for $M=3$ in general and provide one exemplary SM construction.

\vspace{-4mm}

\subsection{Massless spectrum}\label{Ss:Spectrum}

Demanding the absence of matter in the adjoint representation, i.e. {\it rigidity}, for the fractional \mbox{D6-brane stack $a$}
on $T^6/(\Z_2 \times \Z_6 \times \OR)$
with discrete torsion, where the generators $\theta$ and $\omega$ act on the complex coordinates $z_i$ per two-torus $T^2_{(i)}$ as
$\theta^k \omega^l: z_i \to e^{2\pi i (k v_i + l w_i)} z_i
\text{ with }
\vec{v}=\frac{1}{2}(1,-1,0), 
\vec{w}=\frac{1}{6}(0,1,-1)$, 
 amounts to only two allowed bulk orbits, either parallel to the $\OR$-invariant plane or at angle $\frac{\pi}{6}(1,0,-1)$.
The anti-holomorphic involution accompanying the worldsheet parity operator $\Omega$ is here given by
${\cal R}: z_i \to \ov{z}_i$. 
In~\cite{Ecker:2014hma}, we showed that there exist only two physically inequivalent orientations of the $SU(2)^2 \times SU(3) \times SU(3)$ 
background lattice w.r.t. ${\cal R}$, denoted by {\bf aAA} and {\bf bAA}.  
The existence of discrete torsion enforces an odd number of exotic O6-plane orbits $\OR\Z_2^{(k), k\in\{0,1,2,3\}}$~\cite{Blumenhagen:2005tn,Forste:2010gw},
and supersymmetric (SUSY) D6-brane model building  on $T^6/(\Z_2 \times \Z_6 \times \OR)$
can only be achieved for two inequivalent choices \linebreak  $(\eta_{\OR},\eta_{\OR\Z_2^{(1)}},\eta_{\OR\Z_2^{(2)}},\eta_{\OR\Z_2^{(3)}})=$$(-1,1,1,1)$ 
or $(1,1,\underline{1,-1})$. The first choice allows for D6-branes at arbitrary angles $(\phi_1,\phi_2,-\phi_1-\phi_2)$ while the last two choices admit only D6-branes at angles $(0,\phi,-\phi)$.

SUSY bulk orbits leading to spectra without chiral matter in the symmetric representation of $U(3)_a$ - or some GUT version $U(4)_a$ or $U(5)_a$ -
 and at most net-chirality three in the 
antisymmetric sector are constrained to the same kinds of bulk orbits as required for rigidity  with now the additional constraint on the torus wrapping numbers $(n^1_a,m^1_a)=(1,m^1_a)$ with
$m^1_a \in \{1, \ldots, 6 \}_{\bf aAA}$ and $\{0, \ldots, 4\}_{\bf bAA}$ for the non-trivial angle $\frac{\pi}{6}$ on $T^2_{(1)}$.

For any $SU(5)$ GUT, Pati-Salam or SM construction, at least a second stack $b$ with three chiral generations $[(3-z) \times (\N_a,\ov{\N}_b) + z \times (\N_a,\N_b)]$ (with $0 \leqslant z \leqslant 3$)
has to be added, and later on all bulk plus exceptional RR tadpole cancellation conditions and the K-theory constraints need to be satisfied. 
$SU(5)$ GUTs are then ruled out completely, and Pati-Salam models cannot be constructed on the {\bf bAA} lattice, 
while two prototype Pati-Salam models on the {\bf aAA} lattice with a `hidden' $SU(6)$ or $SU(2)$ gauge group were found for $\eta_{\OR\Z_2^{(2 \text{ or } 3)}}=-1$ in~\cite{Ecker:2014hma}.

The systematic scan for `$\varrho$-independent' configuration, i.e. models that are SUSY for arbitrary values of the complex structure parameter $\varrho$ on $T^2_{(1)}$
with all D6-branes at angles $(0,\phi,-\phi)$ and/or $(\frac{\pi}{2},\phi-\frac{\pi}{2},-\phi)$, 
can be extended to left-right symmetric and SM vacua. For example, a SM with full gauge group
$(SU(3)_a \times USp(2)_b \times SU(4)_h)_{U(1)_Y}^{(U(1)_{PQ}),\Z_3}$ can be constructed, see~\cite{Ecker:2015vea} for details on the exact D6-brane configuration. 
 The massless open string spectrum of this model consists of:
  \begin{itemize}
\item 
the {\it chiral} SM-like sector plus some vector-like (w.r.t. the SM group) down-quark and lepton pairs:
{\small
\begin{equation*}
\begin{aligned}
& 3 \times \bigl[  (\3, \2, \1)_{1/6}^{(0),0}  + 2 \times (\ov \3, \1,\1)_{1/3}^{(1),1} + (\3, \1, \1)_{-1/3}^{(1),1} +  (\ov \3, \1, \1)_{-2/3}^{(1),1} \bigr]
\\
&+ 3 \times \bigl[(\1, \2,\1)_{1/2}^{(1),1}  + 2 \times (\1,\2,\1)_{-1/2}^{(1),1} +  (\1,\1,\1)_{0}^{(-2),1} +  (\1,\1,\1)_{1}^{(0),0} \bigr]
\\
& =3 \times \bigl[ Q_L  + 2 \times  d_R + \ov{d_R} +  u_R \bigr] + 3 \times \bigl[ H_u/\ov{L} + 2 \times L +  \nu_R + e_R \bigr]
,
\end{aligned}
\end{equation*}
}
\item  an extended Higgs sector: 
{ \small $3 \times [H_u+H_d] + 2 \times  [ \tilde H_u  + \tilde H_d]=
3 \times [ (\1, \2,\1)_{1/2}^{(1),1} + h.c. ] + 2 \times [ (\1,\2,\1)_{1/2}^{(-1),2}  + h.c.] 
$}
,
\item  QCD axion candidates:  {\small $ 3 \times [\Sigma +  \tilde\Sigma] =   3 \times [ (\1,\1,\1)_{0}^{(-2),1} + h.c. ]$},
\item further vector-like states w.r.t. the SM gauge group: 
{\small
\begin{equation*}
\begin{aligned}
& (5_{\Anti_b} + 4_{\Adj_c} + 5_{\Adj_d}) \times  (\1,\1,\1)_{0}^{(0),0} +3 \times (\1,\2,\4)_{0}^{(0),2} +
\\
& +  6 \times (\1,\1,\ov\4)_{-1/2}^{(-1),0} + 3 \times (\1,\1,\ov\4)_{1/2}^{(-1),0} + 3 \times (\1,\1,\4)_{1/2}^{(-1),1}
\\
&  + \bigl[   2 \times (\3,\1,\ov\4)_{1/6}^{(0),1} + (\3,\1,\4)_{1/6}^{(0),2} + 2\times  ({\bf 3_{A}},\1,\1)_{1/3}^{(0),0} +  h.c. \bigr] 
\\
& + \bigl[ 3 \times (\1,\1,\1)_{1}^{(0),0} +(\1,\1,\1)_{1}^{(-2),1}  + 2 \times (\1,\1,{\bf 6_{A}})_{0}^{(0),1} 
 +  h.c. \bigr] 
 .
\end{aligned}
\end{equation*}
}
\end{itemize}
The upper index in parenthesis corresponds to  the global Peccei-Quinn symmetry at the perturbative level discussed further in section~\ref{Ss:QCDaxion}, and 
the other upper index denotes the $\Z_3$ charge, which remains exact non-perturbatively. 
Since this model does not possess any $(B-L)$-symmetry, left-handed lepton and down-type Higgs multiplets can only be distinguished by their superpotential couplings, 
and QCD axion candidates have the same charges as  right-handed sneutrinos $\tilde{\nu}_R$.

\vspace{-4mm}

\subsection{Yukawa couplings}\label{Ss:Yukawa}

All $m$-point interactions are not only subject to charge selection rules, but due to the existence of $\Z_3$ images of each stack of D6-branes on $T^6/(\Z_2 \times \Z_6 \times \OR)$
the requirement of closed polygons provides an extra `stringy' selection rule. It should also be noted that exact results on Yukawa couplings are only known for the mere six-torus $T^6=(T^2)^3$. 
There, it was argued that Yukawa couplings with all three D6-branes parallel along one two-torus vanish due to an underlying ${\cal N}=2$ SUSY in this sector, which leads to a non-minimal number of zero modes whose integration gives zero~\cite{Cremades:2003qj,Lust:2004cx}. 
In the present case, however, the $\Z_2 \times \Z_2$ subgroup provide a breaking to ${\cal N}=1$ SUSY even though all D6-branes $x \in \{a,b,c,d,h\}$ are at angles $(0,\phi_x,-\phi_x)$, i.e. parallel along $T^2_{(1)}$. We thus expect the CFT result of scattering amplitude computations to be non-vanishing and proceed here by investigating the `stringy' selection rule of closed polygons and their areas.

The up-type Yukawa couplings for example take the form 
\begin{equation*}
{\cal W}_{\text{Yukawa}} \supset y_u^{(ijk)} \,Q_L^{(i)} \cdot \tilde H_u^{(j)} u_R^{(k)},
\end{equation*}
with the discrete $\Z_3$ charge and the perturbatively existent global $U(1)_{PQ}$ charge preventing Yukawa couplings to the other up-Higgses $H_u^{(i)}$.
The worldsheet instanton contribution to the Yukawa coupling scales with the exponential of the areas of the triangles on $T^2_{(2)} \times T^2_{(3)}$, for example:
 \begin{equation*}
  \begin{array}{c|c}
  \text{diagonal} & \text{off-diagonal}
  \\\hline
y_u^{(121)} \sim {\cal O}\left(e^{- \frac{4v_2+v_3}{48} }\right)
&
y_u^{(221)} \sim {\cal O}\left(e^{- \frac{16v_2 + v_3}{48}}\right) 
\\
 y_u^{(313)} \sim {\cal O}\left(e^{- \frac{v_2+4 v_3}{48}}\right)
&
y_u^{(312)} \sim {\cal O}\left(e^{-\frac{v_2+16 v_3}{48} }\right) 
\end{array}
\end{equation*}
Here, $v_i$ denotes the volume of $T^2_{(i)}$.
As one can read off from the complete list in~\cite{Ecker:2015vea}, the third generation of quarks is heavier than the other two if $v_3 < v_2$,
whereas the diagonal terms are dominant if $v_2 < 5 v_3$. As noted in~ \cite{Honecker:2012jd}, an additional rather mild hierarchy might arise from the 
K\"ahler metrics contained in the prefactors of the physical Yukawa couplings.

A mild anisotropy of the two-torus volumes is thus phenomenologically favoured, while the absolute scale of the bulk K\"ahler moduli 
has to provide also the correct order of magnitude of gauge couplings at one-loop level, see e.g.~\cite{Honecker:2012qr} for a discussion on the other 
$T^6/(\Z_2 \times \Z_6' \times \OR)$ orientifold with the $\Z_3 \subset \Z_6'$ acting along the full six-torus $(T^2)^3$.

\vspace{-4mm}

\section{Moduli Stabilisation}\label{S:Moduli}

The closed string sector contains $h_{21}$ complex structure moduli, which for $T^6/(\Z_2 \times \Z_6 \times \OR)$ with discrete torsion stem from the various bulk and twisted sectors as follows,
$\begin{aligned}
h_{21}  &  =
\boxed{h_{21}^{\text{bulk}} + h_{21}^{(\underline{0,\frac{1}{2},-\frac{1}{2}})}  }
+ h_{21}^{(0,\frac{1}{6},-\frac{1}{6})} + h_{21}^{(0,\frac{1}{3},-\frac{1}{3})} 
\\
& =
\boxed{1+ (6 + 2 \times 4)} + 2+2  =19
,
\end{aligned}
$ 

\noindent
but only the bulk and $\Z_2$ twisted sectors highlighted in boxes are accessible for D6-brane model building with known underlying CFT techniques~\cite{Blumenhagen:2006ci,Forste:2010gw}.
By construction, the $\varrho$-independent SM of section~\ref{Ss:Spectrum} has the bulk complex structure on $T^2_{(1)}$ as free parameter, and three massive $U(1)$ factors 
absorb three axionic partners of $\Z_2$ twisted complex structure moduli as longitudinal modes, simultaneously stabilising the corresponding geometric moduli due to SUSY.

More $\Z_2$ twisted moduli can, however, be stabilised due to their couplings to D-branes as noted in~\cite{Blaszczyk:2014xla,Blaszczyk:2015oia}. As can be verified explicitly 
by computing the integral 
$\int_{\Pi_a} \!\! \Omega_3 = \! \int_{\Pi_a} \!\! \frac{dx_1dx_2dx_3}{y}  \! \stackrel{\text{\tiny SUSY}}{=}\!\! \int_{\Pi_a}\!\! \Re(\Omega_3)$
with $(x_i,v_i,y_i)$ projective coordinates in $\mathbb{P}_{112}^2$ in the patch $v_i \equiv 1$ with  $y \equiv y_1y_2y_3$ and $T^6/(\Z_2 \times \Z_2 \times \OR)/\Z_3$ obtained 
by a polynomial constraint (for $\Z_2 \times \Z_2$) and identifications (for $\Z_3$), 
deformations away from the singular orbifold point break SUSY if a D6-brane supporting some $U(1)$ gauge factor couples to them.
From a geometric point of view, this behaviour is encoded in the $\Z_2$ twisted contributions within $\Pi^{\Z_2^{(i)}}_a \neq \OR ( \Pi^{\Z_2^{(i)}}_a)$ of the three-cycle,
\begin{equation*}
\Pi^{\text{frac}}_a = \frac{1}{4} \left(\Pi^{\text{bulk}}_a + \sum_{i=1}^3 \Pi^{\Z_2^{(i)}}_a
\right) \begin{array}{c} \stackrel{SO/USp(2)_a}{=} \\ \stackrel{U(1)_a}{\neq} \end{array} \OR( \Pi^{\text{frac}}_a)
,
\end{equation*}
while from a field theory perspective the periods $\int_{\Pi_a} \!\! \Omega_3$ produce D-terms for $U(1)_a$ gauge factors.
The counting of $\OR$-odd couplings in the SM example of section~\ref{Ss:Spectrum} allows for a partial stabilisation of deformation moduli,
\begin{equation}\label{Eq:Counting-stabilised}
\bigl[h_{21}^{\text{bulk}} + h_{21}^{(\underline{0,\frac{1}{2},-\frac{1}{2}})} \bigr]_{\text{stabilised}} 
\leqslant 0+ (2+2 \times 2)=6
,
\end{equation}
with the others not coupling to any of the D6-branes. 
As noted in~\cite{Blaszczyk:2015oia}, however, if two stacks of D-branes have identical size, such as $U(1)_c \times U(1)_d$ in the SM example at hand, and bifundamental charged scalars exist, 
such as the right-handed sneutrinos and axions, some matter {\it vev} can in principle compensate a defomation {\it vev} located at the same $\Z_2$ fixed point 
such that the breaking $U(1)_c \times U(1)_d \to U(1)_{\text{diag}}$ constitutes a flat direction. 
The counting in~\eqref{Eq:Counting-stabilised} thus constitutes only an upper bound for the two moduli in the $\Z_2^{(2)}$ twisted sector.

The closed string sector of Type IIA/$\OR$ also contains  $h_{11}^+$ vectors and $h_{11}^-$ K\"ahler moduli,
which in the $T^6/(\Z_2 \times \Z_6 \times \OR)$ example of section~\ref{Ss:Spectrum} amount to

\noindent
$\left(\!\!\begin{array}{c} h_{11}^+ \\ h_{11}^-
\end{array}\!\!\right)
\stackrel{\eta_{\OR\Z_2^{(3)}}=-1}{=}
\left(\!\!\begin{array}{c}
h_{11}^{(\frac{1}{2},-\frac{1}{3},-\frac{1}{6})}
\\
h_{11}^{\text{bulk}} + h_{11}^{(\frac{1}{2},-\frac{1}{6},-\frac{1}{3})} + h_{11}^{(0,\frac{1}{3},-\frac{1}{3})} 
\end{array}\!\!\right)
=\left(\!\!\begin{array}{c} 
4 \\
3 +4 + 8
\end{array} \!\!\right)
$,
with the twisted K\"ahler moduli completely decoupled from the SM sector.

\vspace{-4mm}
 
\section{Towards Cosmology}\label{S:Cosmo}

\vspace{-4mm}

\subsection{QCD axion from open strings}\label{Ss:QCDaxion}

The DFSZ model~\cite{Dine:1981rt,Zhitnitsky:1980tq} for the QCD axion can be straightforwardly supersymmetrised.
Its embedding in Type II string theory is, however, constrained by the gauge sector origin from open strings. Since some quarks as well 
as the Higgses are expected to be charged under the global Peccei-Quinn symmetry $U(1)_{PQ}$, which has to be orthogonal to the hypercharge
defined in~\eqref{Eq:hypercharge}, there exist only two possible realisations: $U(1)_{\text{PQ}} \simeq U(1)_b$  or $U(1)_{c - d}$
with the QCD axion realised as $(\1_{\Anti_b})_{\pm 2}$ or $(\1_{cd})_{\pm 2}$~\cite{Honecker:2013mya}. 
The latter occurs in the SM example of section~\ref{Ss:Spectrum}.
Note also that all of the stringy QCD axion candidates necessarily carry twice the charge w.r.t. the original DFSZ model.

The scalar potential can be written as a sum of various contributions, $ V_{\text{scalar}} =  V_F  + V_D  +  V_{\rm soft}$.
For the SM example of section~\ref{Ss:Spectrum}, the Higgs-axion potential arises schematically from
${\cal W}_{DFSZ} =  \mu  \Sigma H_u \cdot \tilde{H}_d + \tilde{\mu}   \tilde{\Sigma} \tilde{H}_u \cdot H_d$
due to the $\Z_3$ selection rule, see~\cite{Ecker:2015vea} for details.
The first kind of axion multiplets $\Sigma$ is also relevant for providing masses to the vector-like down-quark pairs,
${\cal W} \supset \kappa \ov{d}_R  \Sigma d_R$, which requires some tuning of parameters to keep the $\mu$-term small while 
decoupling the vector-like states from the light spectrum.

\vspace{-4mm}

\subsection{Mixing of open and closed string axions}

The complex scalar $\sigma = \frac{v+s(x)}{\sqrt{2}} e^{i \frac{ a(x)}{v}}$ in the multiplet $\Sigma$ contains the open string axion $a$, 
which generically mixes with the closed string axion $\xi$ providing the longitudinal mode of some $U(1)_{\text{massive}}$ gauge boson. 
The mass eigenstates are given by:
\begin{equation*}
 \zeta_{\text{massive}} = \frac{M_{\text{string}}\,  \xi +  q v\,  a}{\sqrt{M_{\text{string}}^2 + q^2 v^2}}, \quad 
 \alpha_{\text{massless}} = \frac{M_{\text{string}}\,  a  -   q v\,  \xi}{\sqrt{M_{\text{string}}^2 +  q^2 v^2}}
 ,
\end{equation*}
and the corresponding axion decay constants read:
\begin{equation*}
f_{\zeta} \simeq \frac{\sqrt{M_{\text{string}}^2 + (qv)^2}}{2}, \quad 
f_{\alpha} \simeq \frac{M_{\text{string}}\, q v \sqrt{M_{\text{string}}^2 + (qv)^2}}{ (M_{\text{string}}^2 - (qv)^2)}
,
\end{equation*}
see~\cite{Honecker:2013mya} for further details.
In the field theory regime $M_{\text{string}} \gg v$, however, the mass eigenstates are very well approximated 
by the closed and open string axions,
$\zeta \simeq  \xi_{\text{closed}}$, $\alpha \simeq  a_{\text{open}}$
with $f_{\zeta}  \simeq  \frac{M_{\text{string}}}{2}$ and $f_{\alpha} \simeq qv$,
which justifies to decouple the discussion of a closed string axion as inflaton from that of the open string QCD axion as solution to the strong CP problem. 

\vspace{-4mm}

\section{Conclusions and Outlook}\label{S:Conclusions}

We presented here a systematic search for new particle physics models on the $T^6(\Z_2 \times \Z_6 \times \OR)$ background with discrete torsion 
and argued that the presence of D6-branes stabilises some complex structure deformation moduli at the orbifold point, while other twisted moduli completely
decouple from the SM sector. We then proceeded to address the origin of the QCD axion from open strings, which might act as a dark sector component, 
and we showed that its field theory can be studied independently from some closed string axion acting as inflaton.

It remains to derive exact field theory results on $m$-point couplings for orbifold backgrounds, which will then allow to scan favourable
 parameter ranges  for gauge and Yukawa couplings as well as axion decay constants.

\noindent
{\bf Acknowledgements:} It is a pleasure to thank M. Blaszczyk, J. Ecker, I. Koltermann and especially W. Staessens for collaboration on various aspects of the project presented here,
and the GGI Florence for hospitality while completing the manuscript.
This work is partially supported by the {\it Cluster of Excellence PRISMA} DFG no. EXC 1098, the DFG research grant HO 4166/2-1,
and the COST Action MP1210.

\end{document}